# The interstellar gas-phase chemistry of HCN and HNC

*Jean-Christophe Loison[1,2]\*, Valentine Wakelam[3,4] and Kevin M. Hickson[1,2]*

\*Corresponding author: jc.loison@ism.u-bordeaux1.fr

[1] *Univ. Bordeaux, ISM, UMR 5255, F-33400 Talence, France*
[2] *CNRS, ISM, UMR 5255, F-33400 Talence, France*
[3] *Univ. Bordeaux, LAB, UMR 5804, F-33270, Floirac, France.*
[4] *CNRS, LAB, UMR 5804, F-33270, Floirac, France*

We review the reactions involving HCN and HNC in dark molecular clouds to elucidate new chemical sources and sinks of these isomers. We find that the most important reactions for the HCN-HNC system are Dissociative Recombination (DR) reactions of $HCNH^+$ ($HCNH^+ + e^-$), the ionic $CN + H_3^+$, $HCN + C^+$, HCN and HNC reactions with $H^+/He^+/H_3^+/H_3O^+/HCO^+$, the N + $CH_2$ reaction and two new reactions: H + CCN and C + HNC. We test the effect of the new rate constants and branching ratios on the predictions of gas-grain chemical models for dark cloud conditions. The rapid C + HNC reaction keeps the HCN/HNC ratio significantly above one as long as the carbon atom abundance remains high. However, the reaction of HCN with $H_3^+$ followed by DR of $HCNH^+$ acts to isomerize HCN into HNC when carbon atoms and CO are depleted leading to a HCN/HNC ratio close to or slightly greater than 1. This agrees well with observations in TMC-1 and L134N taking into consideration the overestimation of HNC abundances through the use of the same rotational excitation rate constants for HNC as for HCN in many radiative transfer models.

## 1 INTRODUCTION

Both of the isomers hydrogen cyanide (HCN) and hydrogen isocyanide (HNC) are ubiquitous in the interstellar medium. They have been detected in diffuse clouds (Liszt & Lucas 2001), in translucent molecular clouds (Turner *et al.* 1997), in dark interstellar clouds (Pratap *et al.* 1997, Hirota *et al.* 1998, Irvine & Schloerb 1984) in starless cores (Hily-Blant *et al.* 2010) and in star forming regions (Jorgensen *et al.* 2004, Godard *et al.* 2010). Although HNC is less stable than HCN by 55 kJ/mol (DePrince III & Mazziotti 2008, Hansel *et al.* 1998, Baulch *et al.* 2005) with an isomerization barrier for passage from HNC to HCN calculated to be equal to 124 kJ/mol (DePrince III & Mazziotti 2008), its abundance is often comparable to that of HCN, especially at temperatures around 10 K where it is extremely thermodynamically unfavored (Hirota et al. 1998, Irvine & Schloerb 1984, Tennekes *et al.* 2006, Sarrasin *et al.* 2010). The HCN/HNC ratio is observed to be much greater than 1 in hot cores (Schilke *et al.* 1992) and in Young Stellar Objects (YSOs) (Schöier *et al.* 2002). Thus it appears likely that competitive rate processes control the HCN/HNC ratio. In this study we performed a thorough review of the various reactions producing and consuming HCN and HNC. We also looked for new sources and sinks of HCN and HNC, considering reactions involving the most abundant species in diffuse and dense molecular clouds ($C^+$, C, N, O and H) and paying particular attention to certain carbon and nitrogen containing species such as $H_2CN$, $CH_2NH$, CCN, HCCN and $H_2CCN$. The most important reactions for the HCN/HNC system are presented in the annex. We found one new efficient HCN production pathway, H + CCN, which is described in a separate paper in preparation, but no new efficient HNC production mechanisms (HNC is produced in the model by the N + $CH_2$ reaction and by the DR of $HCNH^+$). Interestingly, we have introduced the C + HNC → C + HCN reaction which is a very efficient HNC → HCN isomerisation mechanism as long as the atomic carbon abundance remains high as shown in the separate paper dedicated to the chemistry. The most important reactions for the HCN, HNC system are the DR reaction of $HCNH^+$ ($HCNH^+ + e^-$), CN + $H_3^+$, HCN + $C^+$, HCN,HNC + $H^+$/$He^+$/$H_3^+$/$H_3O^+$/$HCO^+$ reactions and three neutral reactions: N + $CH_2$, H + CCN and C + HNC. It is worth noting that as for HCN and HNC, there are two isomers of $HCN^+$/$HNC^+$ and three for $HCNH^+$/$H_2NC^+$/$H_2CN^+$. Their thermochemical values are presented in Table 1 with the corresponding references (some values have been calculated at the MRCI+Q level we performed. The correlation energy being calculated with the internally contracted multireference configuration interaction method, along with the Davidson correction (MRCI+Q) for size-consistency (Werner & Knowles 1988) after complete active space self-consistent field (CASSCF) calculations (Werner & Knowles 1985) using the aug-cc-pVQZ basis set and the MOLPRO 2009 program package). CCSD(T) calculations were also performed using the aug-cc-pVTZ basis set and the MOLPRO 2009 program package. The most stable isomeric form of $HCN^+$ ion is $HNC^+$ as shown Figure 1, $HCN^+$ being metastable. We have tried as much as possible to identify the actual isomer ($HCN^+$ or $HNC^+$) produced in the various chemical reactions. However, in dense molecular clouds, the precise identification of the nature of the $HCN^+/HNC^+$ isomer produced in various reactions is not really important as both ions react quickly with $H_2$ leading to $HCNH^+ + H$. For $HCNH^+/H_2NC^+/H_2CN^+$ we consider only the two most stable isomers, $HNCH^+$, which is the most stable one and the $H_2NC^+$ ion (the isomerization of $H_2NC^+$ toward $HCNH^+$ is calculated to have a barrier from the ground state (Talbi & Herbst 1998)) as $H_2NC^+$ was suspected to produce mainly HNC through DR. However the DR of $H_2NC^+$ is 217 kJ/mol more exothermic than the DR of $HCNH^+$ and considering the lack of any specific information there are no obvious reasons to consider than the very excited (HCN,HNC)** species formed through this DR will not follow the same relaxation mechanism, leading to similar amounts of HCN,HNC,CN as the DR of $HCNH^+$ (see annex).

This paper is organized as follows. In Section 2 we present the impact of the new reactions/rate constants for dark cloud modeling, in Section 3 we analyze the chemistry of HCN and HNC in dense molecular clouds and in Section 4 we compare the simulations with observations in dense dark clouds as well as in other objects. Our conclusions are presented in Section 5. The new reactions and rate coefficients are presented and discussed in the annex.

## 2. RESULTS

### 2.1 The chemical model

The new chemical reactions for HCN and HNC formation and destruction listed in Table 2 have been applied to the chemical model Nautilus (Hersant *et al.* 2009, Semenov *et al.* 2010). The version of Nautilus used for this work is the same as the one used by Loison et al. (2013) and we refer to this paper for a complete description of the model. Briefly, with Nautilus, we can compute the gas-phase and dust icy mantle composition as a function of time taking into account a large number of gas-phase and grain surface reactions and gas-grain interactions. The list of gas-phase reactions is based on the public network kida.uva.2011 (http://kida.obs.u-bordeaux1.fr/models, (Wakelam *et al.* 2012) with updates of the carbon chemistry by Loison et al. (2013), whereas the list of grain surface reactions is similar to (Garrod *et al.* 2007). The final list of reactions is composed of 8584 reactions for 703 species, and will be made public on the KIDA database (http://kida.obs.u-bordeaux1.fr/models).

The chemical composition of both the gas-phase and the grain surfaces is computed as a function of time for a gas and dust temperature of 10 K, a total H density of $2\times10^4$ cm$^{-3}$, a cosmic-ray ionization rate of $1.3\times10^{-17}$ s$^{-1}$ and a visual extinction of 10. All elements are assumed to be initially in atomic form, except for hydrogen which is entirely molecular, with abundances listed in Table 1 of (Hincelin *et al.* 2011), the C/O elemental ratio being equal to 0.7 in this study.

**2.2 Model results**

Figures 2 and 3 show the results for the old (Wakelam et al. 2012, Loison *et al.* 2014) and new networks for a selection of species related to HCN and HNC. Some of these molecules have been observed in the two well-studied dark clouds TMC-1 (CP peak) and L134N (North peak) as listed in Table 3. On the figures, we have superimposed the mean observed abundances. The reported abundances for HCN and HNC in the literature have been deduced from observations using the same rotational excitation rate coefficients for HCN and HNC, equal to the HCN + He rate constants calculated by (Green & Thaddeus 1974). The rotational excitation rate constant with $H_2$ being deduced from the one with He. However it has been shown that this approximation leads to a notable overestimation of the rotational rate constant for HNC (Sarrasin *et al.* 2010, Dumouchel *et al.* 2010, Dumouchel *et al.* 2011). The use of the new rotational rate constant will lead to revised HNC/HCN abundance ratios derived from observations from values >1 to values ≤1 (Dumouchel *et al.* 2010).

In Fig. 2, we plot the simulated abundances as a function of time for HCN and HNC, relative to $H_2$, as well as the HCN/HNC ratio with the old and new rate constants (alongside the comparison with observations in TMC-1 (CP)). The main effect of the new rate constant is a strong decrease of the HNC abundance between $2\times10^3$ yr and $2\times10^5$ yr due to the introduction of the rapid C + HNC → C + HCN reaction leading to an HCN/HNC ratio well above 1, as long as the carbon atom abundance is still high.

In Fig. 3, we plot the simulated abundances as a function of time for various species playing an important role in HCN and HNC chemistry: CN, CCN, $H_2$CN, HNCH$^+$, CH$_3$CN and HC$_3$N. Except for CN and $H_2$CN which are clearly overproduced, the agreement between observed (in TMC-1 (CP)) and modeled abundances is relatively good for times between $2\times10^5$ yr and $4\times10^5$ yr, considering the overestimation of the observed HNC abundance. The effect of the new rate constants is small for all species but CCN (and HNC) after $10^5$yr. It should be noted that the CCN radical, produced by the N + CCH reaction and by the O, N + C$_3$N reactions is predicted to have an abundance around $10^{-9}$ (compared to $H_2$), and thus may be detectable.

## 2.3. HCN, HNC chemistry

The chemistry of HCN and HNC involves various neutral and ionic reactions. The main fluxes of production and loss for both HCN and HNC are shown in Fig. 4. HNC production involves a very simple chemical scheme mainly through the DR of HCNH$^+$ with a large contribution from the N + CH$_2$ reaction before 5×10$^4$ yr (and a minor CH$_3$NH$^+$ + e$^-$ contribution, not represented in Fig. 4). HCN production involves many more reactions in addition to the DR of HCNH$^+$ and the N + CH$_2$ reaction. The reactions H + CCN and C + HNC (with fluxes similar to the DR of HCNH$^+$ until 2-3×10$^5$ yr) and minor sources N + CCH and H$^-$ + CN (not shown on Fig. 4 ). The H$^-$ + CN reaction is thought to produce only HCN + e$^-$ and not HNC as H$^-$ should bind exclusively to the carbon atom. Moreover, as most of the energy of the reaction is transformed into kinetic energy of the departing electron the HCN formed cannot isomerize into HNC. The reactions of HCN and HNC with C$^+$, CH$_3^+$, H$^+$ and He$^+$ are identical in current models (they have the same products and rate constants), and the grain sticking probability for these isomers is also considered to be the same. Before 10$^5$ yr, the high fluxes of C + HNC reaction leads to very high HCN abundance. We can separate the HCN and HNC chemistry into three time intervals, highlighting the complex effect of reactions of HCN and HNC with H$^+$, He$^+$, H$_3^+$, H$_3$O$^+$ and HCO$^+$ followed by the DR of HCNH$^+$.

Before 2×10$^3$ yr, the main HCN and HNC production reactions are N + CH$_2$ (CH$_2$ is produced by the radiative association of C with H$_2$ and by the DR of various CH$_x^+$ species) and the DR of HCNH$^+$. The main destruction process is C$^+$ + HCN, HNC → H + CNC$^+$. HCN isomers are not reformed as CNC$^+$ does not react with H$_2$ (Knight *et al.* 1988) but only with electrons leading to C + CN, CN being quickly consumed through its reactions with N and O atoms. Before 2×10$^3$ yr the DR of HCNH$^+$ is a net source of HCN and HNC as HCNH$^+$ is produced essentially from CCN$^+$ + H$_2$. As a result, the HCN/HNC ratio is greater than 1 at early times as there are many more reactions that produce HCN than HNC. Moreover, the isomerization of HCN → HNC which occurs through the reactions HCN + H$_3$O$^+$/HCO$^+$/H$_3^+$ → HCNH$^+$ + H$_2$O/CO/H$_2$ followed by DR (HCNH$^+$ + e$^-$ → HCN, HNC + H) is not efficient enough to compensate the direct production of HCN.

Between $2\times10^3$ yr and $5\times10^5$ yr the situation is very different. The $C_2N^+$ abundance is low as $C_2N^+$ is produced through the $C_2N + C^+$ reaction and after $2\times10^3$ yr, carbon atoms are predominantly in neutral form leading to a very small flux for the $C_2N + C^+$ reaction. Then $HCNH^+$ is only produced through proton transfer from $H_3O^+$, $HCO^+$, $H_3^+$ to HCN (and HNC) and $HCN^+/HNC^+ + H_2$ reactions. Moreover, over this time period, $HNC^+$ and $HCN^+$ are mainly produced from HCN and HNC through the $H^+ + HCN,HNC \rightarrow H + HNC^+$ reaction and by the reactions $He^+ + HCN,HNC \rightarrow He + H + CN^+$ followed by $CN^+ + H_2 \rightarrow H + HCN^+-HNC^+$. As most of the $HCNH^+$ is produced from HCN and HNC, the DR of $HCNH^+$ simply recycles HCN and HNC but does not act as a real source of these isomers. Additionally, as the production of CN from the DR of $HCNH^+$ is greater than $HCNH^+$ production from sources other than HCN or HNC, the DR of $HCNH^+$ acts as a net loss of HCN and HNC as CN is destroyed by reaction with N and O atoms. To highlight the importance of this sink for HCN and HNC, we show in Fig. 5 the HCN and HNC abundances obtained by setting the branching ratio for CN production of the DR of $HCNH^+$ equal to zero. The effect is to increase the HCN and HNC abundances by almost by a factor of 10 whilst the HCN/HNC ratio tends to unity much more quickly when the carbon atom abundance decreases as the HCN $\leftrightarrow$ HNC isomerization effect of the DR of $HCNH^+$ is more efficient. Between $2\times10^3$ yr and $5\times10^5$ yr the major sources of HCN are the $H + CCN$, $N + CH_2$ reactions with minor contributions from the $H^- + CN$ and $N + CCH$ reactions. The direct HNC production is much lower coming only from $N + CH_2$. The efficiency of the $H + CCN$ reaction to produce HCN and not HNC associated with the efficiency of the $HNC + C \rightarrow HCN + C$ reaction leads to a HCN/HNC ratio greater than 1 and as high as 300. Between $1\times10^5$ yr and $5\times10^5$ yr, the chemical scheme is the same but as the carbon abundance is lower (forming CO and/or being depleted onto grains) the direct HCN (and HNC) production is strongly decreased and proton transfer followed by the DR of $HCNH^+$ partly transform HCN into HNC as well as into CN, which is destroyed by reaction with N and O atoms as N and O atoms deplete later than C atoms. The HCN abundance decreases and HNC increases, as the C + HNC reaction becomes inefficient, producing a HCN/HNC ratio close to 1. The precise HCN/HNC ratio value after $3\times10^5$ yr depends directly on the HCN/HNC product branching ratio of the DR of $HCNH^+$. It's worth noting that the isomerization effect of the DR of $HCNH^+$ cannot compete with the quick C + HNC reaction as long as the carbon atom abundance is high. A low atomic C abundance is required to obtain a ratio of HCN/HNC close to 1, and as the stationary carbon abundance after $3\times10^5$ yr is not completely negligible in the model (see Fig. 4), the C + HNC → C + HCN reaction leads to a HCN/HNC ratio slightly above 1 even if HNC/HCN ratio of the DR of HCNH$^+$ is slightly in favor of HNC (as high as HNC = 40 % and HCN = 27% which is likely to be an upper limiting value for HNC production (see annex)).

We plot on Fig. 5, the HCN and HNC abundances when the N + CH$_2$ and H + CCN rate constants are set to zero. The importance of N + CH$_2$, before $1\times10^5$ yr, as an efficient HCN and HNC source is clearly shown as well as the importance of the H + CCN reaction between $5\times10^4$ yr and $5\times10^5$ yr. It should be noted that setting the N + CH$_2$ rate constant to zero has no effect on the HCN and HNC abundances after $1\times10^5$ yr as CH$_2$ strongly decreases after $5\times10^4$ yr and the proton exchange reactions between HCO$^+$, H$_3$O$^+$, H$_3^+$ and HCN or HNC followed by the DR of HCNH$^+$ are quick enough to reach the stationary equilibrium controlled by the CN + O and CN + N reactions. This is not the case for the H + CCN reaction as the CCN abundance remains elevated after $5\times10^4$ yr.

After $5\times10^5$ yr, the abundance of atoms is low because they have already reacted or have depleted onto grains, and CO is also depleted on grains as shown on lower panels of Fig. 2. Then, the chemistry is no longer driven by C, N and O reactions and as CO is also partially depleted, the CO + H$_3^+$ reaction is less efficient, leading to an increase of the H$_3^+$ abundance. However it should be noted that even if atoms have already reacted or have depleted onto grains, high gas phase abundances of CO and N$_2$ lead to non-negligible C$^+$, O and N abundances through reactions with He$^+$. CH$_4$ and NH$_3$ are formed primarily on grains by surface chemistry and are released back to the gas-phase through various non-thermal desorption mechanisms, as described in (Garrod et al. 2007) leading to relatively high abundance as shown on lower panels of Fig.2. These molecules react with H$_3^+$ leading to a relatively rich ionic chemistry and the formation of NH$_2$ and CH$_2$ through DR reactions of NH$_4^+$ and CH$_5^+$, and then to atomic carbon through the CH$_2$ + H and CH + H reactions. As a result of the low atomic abundances and the high H$_3^+$ abundance, the main CN destruction reaction is CN + H$_3^+$ → HCN$^+$ + H$_2$, followed by HCN$^+$ + H$_2$ → H + HCNH$^+$. The DR of HCNH$^+$ is no longer a major loss of HCN and HNC leading mainly to recycling of HCN and HNC and the HCNH$^+$/HCN$^+$/HNC$^+$/HCN/HNC/CN network is almost a closed system. The main losses of HCN-HNC/HNCH$^+$/CN are through the depletion of HCN, HNC and CN onto grains, dissociation through reactions with He$^+$ or Cosmic Rays and CN reactions with O, N, O$_2$, C$_2$H$_2$ and C$_4$H$_2$ (but not with NH$_3$ which leads to HCN), reactions involving relatively low fluxes. The low fluxes for HCN, HNC and CN loss are compensated by direct CN, HCN and

HNC production, involving also relatively low fluxes, through the H + CCN, C + NH$_2$, N + C$_2$, C + NO and H + H$_2$CN reactions. The HNC/HCN ratio is close to 1 due to the efficient recycling of HCN into HNC through the DR of HCNH$^+$, the residual C + HNC → C + HNC flux slightly favoring HCN. To highlight the importance of the CN + H$_3^+$ reaction we plot in Fig. 5 the HCN and HNC abundances when the rate constants of these reaction are set to zero. Turning off the CN + H$_3^+$ reaction leads to a notable decrease of HCN and HNC abundance after 5×10$^5$ yr as the HCNH$^+$/HCN$^+$/HNC$^+$/HCN/HNC/CN network is no longer a closed system and CN is lost through various reactions such CN + O, CN + N, CN + O$_2$.

## 3. COMPARISON WITH OBSERVATIONS

In many regions of the interstellar medium (ISM), and in dark clouds particularly, the HCN and HNC rotational lines are optically thick so that HCN and HNC abundances are inferred from the corresponding H$^{13}$CN and HN$^{13}$C isotopomers. A $^{12}$C/$^{13}$C abundance ratio of between 64 and 68 is usually assumed, although this value has varied over time, with the most recent estimate being 68 (Milam *et al.* 2005). The extrapolation of H$^{13}$CN and HN$^{13}$C relative abundances to those of the major isotopomeric forms are reliable only if HCN and HNC do not undergo significant carbon fractionation. There are some indirect indications that HCN and HNC do not show major carbon fractionation. First, in diffuse clouds where the HCN and HNC chemistry is thought to involve similar reactions to those in dark clouds, (Lucas & Liszt 1998) find a $^{12}$C/$^{13}$C ratio in general equal to 59 ± 2, except toward 3C111(B0415+379) where the H$^{12}$CN/H$^{13}$CN ratio was equal to 170 ± 50. Secondly (Milam et al. 2005) found an average local ISM $^{12}$CN/$^{13}$CN ratio equal to 68 ± 15, although these observations were relatively scattered. These two indications do not constitute definitive proof that carbon fractionation does not occur. Indeed the reactions involving atomic carbon in dark clouds are characterized by higher fluxes than in diffuse clouds where the elemental carbon is mainly in the form of C$^+$. One particularly important fractionation reaction in dark clouds could be the $^{13}$C + H$^{12}$CN reaction. We performed theoretical calculations at various levels (MRCI+C/aug-cc-pVTZ, CCSD(T)/aug-cc-pVTZ and DFT(M06-2X)/cc-pVTZ) showing no barrier in the entrance valley of the $^3$C + $^1$HCN reaction leading to the $^3$HCNH radical, which evolves through cyclic c-HCCN towards the HCCN radical. These results are in good agreement with the calculations of (Mebel & Kaiser 2002) even if they found that the first step of the C + HCN reaction is the formation of cyclic c-HCCN without a barrier. As this reaction does not have any exothermic bimolecular exit channels it could favor carbon exchange as $^{12}C + H^{13}CN$ is 48.4 K lower in energy than $^{13}C + H^{12}CN$. Preliminary calculations show that carbon exchange involves a complex mechanism through a cyclic HNCC intermediate with transition states located close to the energy of the reactants. Then the carbon exchange will be kinetically slow but may be non-negligible inducing some carbon fractionation. Moreover, the average local ISM $^{12}CN/^{13}CN$ ratio equal to 68 ± 15 from (Milam et al. 2005) shows considerable dispersion (with values of $^{12}CN/^{13}CN$ ranging from 18 to 134) and even if CN chemistry is related to the HCN and HNC ones, there are many others sources of CN in addition to the DR of $HCNH^+$ such as N + CH, N + $C_2$, N + $C_2N$, O + $C_2N$, C + $C_2N$, C + NO so that the $H^{12}CN/H^{13}CN$ and $HN^{12}C/HN^{13}C$ ratio may be notably different from the $^{12}CN/^{13}CN$ one. Indeed, $^{13}C$ fractionation could also be different for HCN than for HNC.

The HCN and HNC detection in TMC-1 by (Pratap et al. 1997) show that even for $H^{13}CN$ and $HN^{13}C$, the lines are not optically thin and self-absorption plays a role. In their study, they used the rotational excitation rate constants for HCN with He calculated by Green & Thaddeus (1974) to deduce the rate constants for the rotational excitation of HCN by $H_2$, assuming that the values for HNC-He and HNC-$H_2$ collisions were identical to the HCN ones. (Hirota et al. 1998) performed detailed HCN and HNC observations in various dense molecular clouds leading to similar results for TMC-1 to (Pratap et al. 1997) despite the lower abundances derived from Hirota et al.. Taking into account the various detection uncertainties, the simulated HCN abundance is compatible with a cloud age between $3 \times 10^5$ yr and $6 \times 10^5$ yr varying from $(2-6) \times 10^{-8}$ relative to $H_2$. This value compares with the observed HCN abundance of between $7.1 \times 10^{-9}$ to $3.4 \times 10^{-8}$ relative to $H_2$ (Pratap et al. 1997, Hirota et al. 1998, Irvine & Schloerb 1984). As HCN is a quite unreactive molecule there is not many species directly correlated to it, more precisely there are the HNC, $HCNH^+$ and $CH_3CN$ species. The observed HNC abundance is overestimated due to the underestimation of the rotational excitation rate constants for HNC-$H_2$ collisions, but the agreement between observed and calculated values (see Figure 2) being likely similar that in the case of HCN for a cloud age between $3 \times 10^5$ yr and $6 \times 10^5$ yr. It should be noted that a high HNC abundance with a HCN/HNC ratio close to 1 is clearly indicative of low atomic carbon abundances corresponding to a cloud age above $10^5$ yr in the current model predictions. The calculated $HCNH^+$ abundance, specie directed related to HCN and HNC, fits the observational one for a cloud age equal to $2.5 \times 10^5$ yr (see Figure 3). This good agreement is due to the fact that we chose, among the extrapolated values at 10 K from various experimental measurements between 160 K and 1000K (Adams & Smith 1988, Semaniak *et al.* 2001, McLain & Adams 2009), the rate constant for the HCNH$^+$ + e$^-$ reaction which yielded the best fit to observations. The agreement between the calculated CH$_3$CN abundance and the observed one for TMC-1 is very good, for a cloud age of 3×10$^5$ yr (see Figure 3), but might be fortuitous. Indeed the main source of CH$_3$CN in our model is:

$$HCN + CH_3^+ \rightarrow CH_3CNH^+ + h\nu$$
$$CH_3CNH^+ + e^- \rightarrow CH_3CN + H$$
$$\rightarrow H_2CCN + H + H$$

and these reactions and are not well constrained. The HCN + CH$_3^+$ → CH$_3$CNH$^+$ + hν radiative association rate constant has been deduced at 300K from low pressure experiments (McEwan *et al.* 1980) and the value at 10 K, deduced from statistical calculations (Bass *et al.* 1981, Bates 1983, Herbst 1985) is subject to relatively large uncertainties. Moreover the branching ratios for DR of CH$_3$CNH$^+$ are unknown. Taking into account the uncertainties of the observed values for HCN and HNC, associated to the fact that many critical rate constant are unknown at 10K leading to large uncertainties of the calculated values, we can consider that observations and calculations are in fairly good agreement for a TMC-1 cloud age between 3×10$^5$ yr and 4×10$^5$ yr.

Observations by (Swade 1989) of L134N(L183) concern only the H$^{13}$CN (J=1-0) and HN$^{13}$C (J=1-0) transitions, the lines of the main isotopomers being very optically thick. The abundances relative to H$_2$ are directly deduced from line intensities (assuming local thermodynamic equilibrium conditions, LTE) and the HCN and HNC abundances are deduced from $^{13}$C isotopomers abundance using a $^{12}$C/$^{13}$C ratio equal to 65. The observations of (Dickens *et al.* 2000) also use H$^{13}$CN and HN$^{13}$C as a proxy and they compute the abundances from line intensities using a statistical equilibrium model using the same collisional rates for HNC and HCN, which is known to lead to an overestimation of HNC abundances (Sarrasin et al. 2010, Dumouchel et al. 2010, Dumouchel *et al.* 2011). (Dickens et al. 2000) found higher HCN and HNC densities attributing their differences with (Swade 1989) to the optical depth estimation as Swade assumed optically thin emission, potentially underestimating the column densities. (Hirota et al. 1998) were not able to evaluate the HNC abundance and therefore only reported HCN abundances. Recently (Hily-Blant et al. 2010) determined CN, HCN and HNC abundances in various starless cores. Using an LTE analysis on H$^{13}$CN and H$^{13}$NC isotopomers, the authors found much smaller HCN and HNC abundances in dark cloud L134N than earlier observations. All observations lead to smaller relative abundances of HCN in L134N than in TMC-1 which is compatible with a cloud age above $2\times10^5$ yr in the current model predictions and L134N is considered to be more evolved then TMC-1. The HCN/HNC abundance ratio is always above 1 and as high as 3. The determination of the HCN/HNC ratio using non-LTE analyses are likely to be unreliable due to the use of incorrect collisional excitation rates for HNC (Sarrasin et al. 2010, Dumouchel et al. 2010, Dumouchel et al. 2011). Taking into account the new collisional excitation rates will lead to smaller HNC abundances, almost certainly leading to a HCN/HNC ratio above 1 in agreement with our simulations. The results from LTE analyses are somewhat surprising as they lead to similar HCN/HNC ratios as the non-LTE treatment using incorrect collisional excitation rate constants. That may indicate that either the transitions used are optically thick or thermal equilibrium is not reached.

In the Young Stellar Object (YSO) IRAS16293-2422, the HCN/HNC ratio derived from optically thin $H^{13}CN$ and $HN^{13}C$ lines leads to a ratio of $H^{13}CN/HN^{13}C$ of 3.5 (Schöier et al. 2002). Chemical models by (Bruderer *et al.* 2009a, Bruderer *et al.* 2009b) of the outer envelope of IRAS16293, with a low temperature similar to dense clouds and using the UMIST chemical network, predict a HCN/HNC abundance ratio of 1 in disagreement with the observations (Schöier et al. 2002). Since the observed abundance of $HCO^+$ in IRAS16293 is high ($1.4\times10^{-9}$ compared to $H_2$, (Schöier et al. 2002), the $HCN + HCO^+ \rightarrow HCNH^+ + CO$ reaction followed by DR of $HCNH^+$ should be a very efficient way to form HNC from HCN, leading to a HCN/HNC ratio close to 1. Considering the cold chemistry, the only efficient reaction to selectively reconvert HNC to HCN is the $C + HNC \rightarrow C + HCN$ reaction.

In higher temperature regions, the chemistry may be different. For example the $CN + H_2$ reaction showing an activation barrier of 2370 K (Jacobs *et al.* 1989) is an efficient source of HCN alone. The main HNC production is then only the DR of $HCNH^+$ after proton exchange from $HCO^+$ and $H_3^+$ to HCN. Moreover, if the O + HCN reaction shows a relatively high barrier close to 4000 K (Sander *et al.* 2011) and if the H + HCN reaction also shows a relatively high barrier without any exothermic exit channels (Talbi & Ellinger 1996, Sumathi & Nguyen 1998, Petrie 2002), the corresponding HNC reactions are significantly faster. The $H + HNC \rightarrow H + HCN$ reaction has been calculated to possess a barrier estimated between 800 K and 1400 K (Talbi & Ellinger 1996, Sumathi & Nguyen 1998, Petrie 2002) and the barrier for the O + HNC is calculated to be equal to 1400 K only at the M06-2X/cc-pVTZ level (this work). Then, the O and H reactions may become an efficient HNC destruction pathway at higher temperatures.

# 4. CONCLUSION

In this study, we have reviewed the gas-phase reactions controlling the HCN/HNC abundance ratio in dense cloud conditions. We show that the low temperature chemistry of HCN and HNC is dominated by the DR of $HCNH^+$ ($HCNH^+ + e^-$), the ionic $CN + H_3^+$, $HCN + C^+$, $HCN, HNC + H^+$, $He^+$, $H_3^+$, $H_3O^+$, $HCO^+$ reactions as well as the neutral $N + CH_2$, $H + CCN$ and $C + HNC$ reactions. The introduction of the $C + HNC$ reaction prevents the HCN/HNC ratio from reaching unity as long as atomic carbon remains in the gas-phase. Then the observation in dark molecular clouds of a ratio close to 1 is a strong indication that carbon atoms are mostly either in the form of CO or depleted onto interstellar grains in these objects. We highlight the complex effect of the DR of $HCNH^+$ showing that between $2 \times 10^3$ yr and $5 \times 10^5$ yr the DR of $HCNH^+$ acts mainly as a sink for HCN through the $CN + H + H$ channel as CN is mainly consumed by reaction with O and N atoms. After $5 \times 10^5$ yr, the DR of $HCNH^+$ acts mainly as an $HCN \leftrightarrow HNC$ isomerization mechanism as the abundance of atoms is low so that CN is no longer consumed by reactions with O and N atoms but mainly with $H_3^+$ leading to $HNC^+$ (and also likely $HCN^+$) which react with $H_2$ leading to $HCNH^+$. Then even if HCN production is more important than HNC production by neutral reactions, the HCN is isomerized into HNC through the $HCN + H_3^+ \rightarrow HCNH^+ + H_2$ reaction followed by $HCNH^+ + e^- \rightarrow HCN, HNC + H$, the CN produced by $HCNH^+ + e^-$ being quickly transformed back to $HCNH^+$. The precise HCN/HNC ratio value is directly dependent on the HCN/HNC product branching ratio of the DR of $HCNH^+$ which should therefore be close to 1 after $3 \times 10^5$ yr.

The authors thank the French CNRS/INSU program PCMI for their partial support of this work. VW and FH research are funded by the ERC Starting Grant (3DICE, grant agreement 336474).

**Annex: Chemistry Review**

**New potential HCN and HNC sources:**

The $H_2CN + H$ reaction is an efficient source of HCN and HNC as shown in a recent article (Hébrard *et al.* 2012). However the $H_2CN$, produced mainly by the $N + CH_3$ reaction (Marston *et al.* 1989a, Marston *et al.* 1989b), has a low abundance in simulations and observations (Ohishi *et al.* 1994) leading to a minor role for HCN and HNC production in interstellar media.

The $CH_2NH + C$ reaction may be a source of HCN and HNC but the production of $CH_2NH$ (through DR of $CH_3NH_3^+$ and $CH_3NH_2^+$ with a small contribution from the $CH + NH_3$ reaction (Bocherel *et al.* 1996, Blitz *et al.* 2012, Zabarnick *et al.* 1989)) is not efficient in molecular clouds leading to a low simulated abundance and only minor production of HCN and HNC.

The CCN radical could be a major source of HCN through the $CCN + H$ reaction as well as of CN production through the CCN + N, O and C reactions although the CCN radical itself has not yet been detected in interstellar media. CCN is thought to be produced by the reactions $N + C_2H$ and $O + C_3N$ and by the DR reactions of $CH_2CN^+$ and $HCCN^+$. The chemical network of CCN has been recently updated (ref). The isomeric form CNC does not have a permanent dipole so it is not detectable by microwave spectroscopy. Moreover, it is likely to be much less abundant than CCN, even if it is more stable, as there are fewer production pathways. As its reactivity toward O, N and C atoms should be similar to the CCN radical, we merged this isomer with the CCN one.

HCCN is not present in KIDA or in other astrochemical networks, but is present in atmospheric models of Titan (Hébrard et al. 2012, Lavvas *et al.* 2008) where it is suspected to play an important role. However, it has not been detected in either TMC-1 or L-134N, with an upper limit in TMC-1 equal to $2 \times 10^{-10}$ with respect to molecular hydrogen (McGonagle & Irvine 1996). It may be produced by the CH + HCN, CH + HNC, $C + H_2CN$ reactions and by DR of $CH_2CN^+$ and $CH_3CN^+$ involving relatively small fluxes. Moreover, HCCN is supposed to react quickly with H (Osamura & Petrie 2004, Takayanagi *et al.* 1998) and very likely also with C, N and O atoms without producing HCN or HNC. We neglect this species in this study and did not introduce it into KIDA.

The $H_2CCN$ radical is already present in the KIDA database and has been detected in TMC-1 (Irvine & et al. 1988). It is thought to be to be produced mainly

- through Dissociative Recombination with electrons (DR) of $CH_3CNH^+$ ($CH_3CNH^+$ being produced through the $HCN + CH_3^+$ radiative association re),

- through DR of $CH_3CN^+$ and through the $CH_3CN^+ + CO \rightarrow HCO^+ + H_2CCN$ reaction ($CH_3CN^+$ being produced through the $N + C_2H_4^+ \rightarrow CH_3CN^+ + H$ reaction (Scott *et al.* 1999)),

- and through the neutral $N + C_2H_3$ reaction.

These species, and most of the reactions, were already present in KIDA but we modify some of the rate constants or branching ratios according to those given in previous work (refs). In particular we complete the neutral chemical network by adding the reactions of $H_2CCN$ with atoms which induces some changes for the $H_2CCN$ abundance but not for HCN and HNC.

**$HCNH^+, H_2NC^+ + e^-$:**

The DR reaction of $HCNH^+$ is of crucial importance for HCN and HNC chemistry. DR of $HCNH^+$ leads to three sets of products:

$^1HCNH^+ + e^- \quad \rightarrow\ ^2H +\ ^1HCN \qquad \Delta H_r(298) = -596$ kJ/mol
$\qquad\qquad\qquad \rightarrow\ ^2H +\ ^1HNC \qquad \Delta H_r(298) = -543$ kJ/mol
$\qquad\qquad\qquad \rightarrow\ ^2H +\ ^2H +\ ^2CN \qquad \Delta H_r(298) = -78$ kJ/mol

The reaction rate constant has been measured three times between 160 K and 1000K (Adams & Smith 1988, Semaniak *et al.* 2001, McLain & Adams 2009) leading to a very high extrapolated value at 10K (between $2\times10^{-6}$ to $3\times10^{-5}$ from ion-molecule rate constant theory). A precise value of the DR rate constant is not critical for HCN and HNC abundances as it is the main sink for $HCNH^+$ which controls the stationary $HCNH^+$ abundance. We chose the (Semaniak et al. 2001) values leading to the best agreement between our $HCNH^+$ abundance with observations in TMC-1 (CP) (Schilke *et al.* 1991). The total internal energy of the excited intermediate HCNH is 672 kJ/mol (Jursic 1999, Nesbitt *et al.* 1991), see also Table 1. There are various theoretical studies of the DR reaction of $HCNH^+$ leading to similar results despite disagreement for the mechanism which involves either a direct formation pathway (Talbi & Ellinger 1998, Hickman *et al.* 2005) or an indirect process (Shiba *et al.* 1998, Taketsugu *et al.* 2004, Ishii *et al.* 2006) as these authors found no possibility for the direct process. However all studies of the direct and indirect processes lead to similar branching ratios for HCN and HNC formation. There are few scattered experimental determinations of the product branching ratios. (Semaniak et al. 2001) obtained the branching ratios using an ion storage ring. According to their results, the pathway that leads to the combined formation of HCN and HNC accounts for 67% of the total with the pathway leading to H + H + CN accounting for the rest. (Amano *et al.* 2006) using submillimeter-wave transitions, observed HNC and HCN in an extended negative glow discharge of a gas mixture of $CH_4$ and $N_2$. They measured a branching ratio HCN/HNC equal to 3.0 ± 0.6 for the normal species and 1.5 for the deuterated species. However the kinetic conditions in these experiments were not well controlled. In particular, they did not measure the various radicals present in their experiment which could have led to strong deviations as HNC reacts differently from HCN with the radicals and atoms present.

The main part of the exothermic energy released in the $HCNH^+ + e^-$ reaction is carried away as internal energy of the HCN-HNC products as shown by recent experiments (Mendes *et al.* 2012), in good agreement with the fact that 33% of HCN-HNC produced dissociates into CN +H. In interstellar clouds the only way to relax this internal energy is through radiative emission of an infrared photon. As noted by Herbst et al (Herbst *et al.* 2000) the typical time-scales for the HNC-HCN system are $<10^{-13}$ s for interconversion and $<10^{-2}$ s for relaxation by one infrared photon. Thus, as relaxation occurs slowly, isomerization leads to equilibrated isomeric abundances at each internal energy. The final balance is determined at or near the effective barrier to isomerization, which corresponds to the energy of the transition state with rotational quantum number J. In the statistical (RRKM) limit, the final ratio of HNC to HCN varies between 0.85 and 1.3 depending on the value of the angular momentum of the HCN/HNC fragment (angular momentum driven by the loss of an H atom from HCNH) (Herbst et al. 2000). Barger et al (Barger *et al.* 2003) performed full dimensional ab initio calculations of the radiative relaxation of HCN-HNC at energies above the barrier. Their preliminary results show that infrared stabilisation leads to a HNC/HCN ratio close to but less than 1, even if the final branching ratio may be substantially different from unity. In the DR reaction of $HCNH^+$, the branching ratios for HCN and HNC formation are likely to be close to 33% each, but values in the range 27% - 40% are also possible. We tested these various possibilities which lead to only small differences and a HCN/HNC ratio always slightly greater than 1 due to the competition with the C + HNC → C + HCN reaction.

The DR reaction of $H_2NC^+$ is assumed to be similar to the DR reaction of $HCNH^+$, leading initially to highly excited HNC/HCN and an H atom. As the DR reaction of $H_2NC^+$ is 219 kJ/mol more exoenergetic than the DR reaction of $HCNH^+$, so that the HCN-HNC produced should be more excited than from the corresponding $HCNH^+$ reaction. Further evolution of the HCN-HNC produced should be then similar to the case of the DR reaction of $HCNH^+$ and should lead to H + CN , HCN and HNC. If, as for $HCNH^+ + e^-$ reaction, the main part of the exothermicity of the DR reaction is carried away as internal energy of the HCN-

HNC products (Mendes et al. 2012), the CN + H + H branching ratio may be even larger for DR of $H_2NC^+$ than for DR of $HCNH^+$. The DR reaction of $H_2CN^+$ plays only a small role in our model so we use the same branching ratios as for the DR reaction of $HCNH^+$.

**HCN, HNC + $C^+$ and $CNC^+/CCN^+$ + $H_2$ reactions:**

The reactions of $C^+$ with HCN and HNC are critical reactions for the HCN-HNC system. Indeed, if the HNC + $C^+$ reaction can lead to two sets of products $CCN^+$ + H and $CNC^+$ + H, the HCN + $C^+$ reaction leads only to $CNC^+$ + H products as $CCN^+$ + H is endothermic (Knight et al. 1988). $CCN^+$ reacts with $H_2$ leading to $HCNH^+$ + C and then back to HCN and HNC through DR of $HCNH^+$. However there is no exothermic channel for the $CNC^+$ + $H_2$ reaction. As a result, $CNC^+$ mainly reacts with electrons leading to CN + C and potentially $C_2$ + N. As the main loss of CN is through reaction with O and N atoms before $5 \times 10^5$ yr, $CNC^+$ ions do not lead back to HCN or HNC so that the HCN/HNC + $C^+$ → $CNC^+$ + H reactions are efficient HCN destruction pathways when O and N atom abundances are high.

**N + $CH_3^+$:**

This reaction has been studied theoretically (Talbi 1999b) as well as experimentally at room temperature (Scott et al. 1999). There is no barrier in the entrance valley leading to an adduct $^4NCH_3^+$, 92 kJ/mol below the entrance level. There are no spin allowed exothermic products for the $^4N + {}^1CH_3^+$ reaction but many ion-atom reactions do not appear to conserve spin (Ferguson 1983). The $^4A"$ adduct is likely to be coupled to the $^2A'$ state through spin orbit coupling leading to the production of $H_2CN^+$ (or $HCNH^+$) + H as well as minor products $HCN^+$ ($HNC^+$) + $H_2$. Considering that the adduct is high in energy (although lower in energy than the reactants), its lifetime with respect to back dissociation is likely to be short leading to a low intersystem crossing rate and a relatively low total rate constant. The precise identification of product isomers is not critical as the main $H_2CN^+$ and $HCNH^+$ reaction is with electrons leading in both cases to similar products. As $HCN^+$ and $HNC^+$ react quickly with $H_2$ both of these ions lead to $HCNH^+$ formation.

**N + $CH_2$:**

There is no barrier for this reaction as shown by ab-initio calculations (Talbi 1999a, Herbst et al. 2000, Takahashi & Takayanagi 2007). The ground electronic state $^4N$ is singly degenerate and $^4N + {}^3CH_2$ reactants correlate adiabatically with sextet, quadruplet and doublet states leading exothermically to products. The products HCN/HNC + H in their ground state correlate only to doublet states so there is an electronic degeneracy factor equal to 1/6 leading to a capture rate constant equal to $8.0 \times 10^{-11} \times (T/300)^{0.17}$ cm$^3$ molecule$^{-1}$ s$^{-1}$ (Herbst et al. 2000). It should be noted that the quadruplet surface also presents no barrier in the entrance valley and may participate to the reaction (leading to $^3$HCN/HNC + H), so that the rate constant may be substantially higher, up to $2.4 \times 10^{-10} \times (T/300)^{0.17}$ cm$^3$ molecule$^{-1}$ s$^{-1}$ (Herbst et al. 2000). The main products from ab-initio calculations are HCN + H. Nevertheless, as the isomerization energy of the HCN $\rightarrow$ HNC conversion is only 186 kJ/mol (DePrince III & Mazziotti 2008), there is so much excess energy available in this reaction (510 kJ/mol) that the HCN product undergoes efficient isomerization leading to near equal production rates of the two isomers. A precise study of the internal energy distribution is necessary to get an accurate branching ratio, although the HCN is likely to be slightly more abundant that HNC as all the HCN produced with less than 186 kJ/mol of excess energy will be unable to isomerize. As the final branching ratio will be in favor of HCN, we propose a ratio HCN/HNC = 5/3.

**N + C$_2$H / H + C$_2$N / C + HNC:**

These reactions have been studied by a coupled ab-initio/RRKM study presented in a separate paper. Only summaries are presented here.

**N + C$_2$H:**

The N($^4$S) + C$_2$H(X$^2\Sigma^+$) reagents correlate with one quintet surface and one triplet surface, the triplet surface does not show any barrier in the entrance valley at the MRCI+Q/aug-cc-pVTZ and CCSD(T)/aug-cc-pVQZ levels, in good agreement with (Takahashi & Takayanagi 2006, Mebel & Kaiser 2002) (it should be noted that the first low lying $^2\Pi$ exited state interferes with the ground X$^2\Sigma$ state in C$_s$ symmetry leading to complex surfaces which likely require multiconfigurational methods of calculation). RRKM calculations lead to 94% of C$_2$N + H (including minor CNC + H channel), 4 % of HCN + C and 2% of HNC + C. Using capture rate theory and considering the degeneracy factor we calculate the rate coefficient to be $k_{N+C2H} = 2 \times 10^{-10} \times (T/300)^{0.17}$ cm$^3$ molecule$^{-1}$ s$^{-1}$ close to the experimental N + CH reaction rate constant (Brownsword *et al.* 1996, Daranlot *et al.* 2013). As the exothermicity is close to the energy of the isomerization barrier, subsequent isomerisation between HCN and HNC is unlikely to occur.

**H + C$_2$N:**

The $^2$H + $^2$CCN reagents correlate with two triplet and two singlet surfaces ($^2$A'⊗($^2$A'+$^2$A") = $^{1,3}$A'+ $^{1,3}$A"). We performed DTF(M06-2X)/cc-pVQZ, MRCI+Q/aug-cc-pVTZ and CCSD(T)/aug-cc-pVQZ calculations showing no barrier for this reaction on the $^3$A" surface, in good agreement with (Mebel & Kaiser 2002, Takahashi & Takayanagi 2006) as well as no barrier on the $^1$A' surface, but the calculations show a high barrier on the $^3$A' surface and a small barrier on the $^1$A" surface. The triplet surface correlates with $^3$C + $^1$HNC, $^3$C + $^1$HCN and $^2$H + $^2$CNC exit channels but the singlet surface only correlates with the $^2$H + $^2$CNC exit channel. RRKM calculations lead to 20 % of CNC + H and 80 % of C + HCN. As we do not consider CNC in this study, we include the production flux of CNC + H production in the back-dissociation of the HCCN adduct toward H + CCN. The global rate constant has been estimated from capture rate theory leading to $k_{H+C2N}$ = 2×10$^{-10}$ cm$^3$ molecule$^{-1}$ s$^{-1}$ in the 10-300K range It should be noted that the uncertainty on the rate constant at 10 K is large.

**C + HNC**

The C($^3$P) + HNC(X$^1\Sigma^+$) reagents correlate with three triplet surfaces. Among the three surfaces, the $^3$A$_2$ surface shows no barrier and the $^3$B$_1$ and $^3$B$_1$ surfaces are repulsive at the MRCI+Q/aug-cc-pVTZ and CCSD(T)/aug-cc-pVQZ levels. Our results are in good agreement with previous calculations at the CASPT2 level (Takahashi & Takayanagi 2006). It is very likely that this important reaction for the HCN/HNC ratio in interstellar media is fast, even at low temperature. We calculate the branching ratio using RRKM theory, leading mainly to C + HCN product with a small contribution from H + CNC, and the global rate constant has been calculated using capture theory leading to $k_{C+HCN}$ = 2×10$^{-10}$ cm$^3$ molecule$^{-1}$ s$^{-1}$ in the 10-300 K range.

**Table 1: Enthalpies of formation of the various isomers**

| Species | $\Delta H_f^{298}$ kJ/mol | reference |
|---|---|---|
| HCN($^1\Sigma^+$) | 135 | (Baulch et al. 2005) |
| HNC($^1\Sigma^+$) | 188 | (Hansel et al. 1998) |
| HCN$^+$($^2\Pi$) | 1447 | (Baulch et al. 2005, Lias *et al.* 1988) |
|  | 1440 | This work (MRCI+Q/cc-pVQZ including ZPE) |
| HNC$^+$($^2\Sigma^+$) | 1336 | (Kraemer *et al.* 1992) |
|  | 1349 | This work (MRCI+Q/cc-pVQZ including ZPE) |
| HCNH$^+$($^1\Sigma^+$) | 949 | (Nesbitt et al. 1991, Jursic 1999) |
| H$_2$NC$^+$($^1A_1$) | 1168 | (Jursic 1999, Talbi & Herbst 1998) |
| H$_2$CN$^+$($^1A_1$) | 1255 | (Jursic 1999, Nesbitt et al. 1991, Talbi 1999b) |
| H$_2$CN($^2B_2$) | 248 | (Zhou & Schlegel 2009, Jursic 1999) |
| HCNH($^2B_2$) | 277 | (Jursic 1999) |
| H$_2$NC($^2B_2$) | 373 | (Jursic 1999) |

**Table 2: Summary of reactions review.**

Temperature range is 10-300K. Definition of α, β, γ, $F_0$, g, Ionpol1 and Ionpol2 can been found in (Wakelam *et al.* 2012, Wakelam *et al.* 2010).

| Reaction | | ΔE kJ/mol | α | β | γ | $F_0$ | g | Ref |
|---|---|---|---|---|---|---|---|---|
| H + H$_2$CN | → HCN + H$_2$ | -331 | 6.0e-11 | 0 | 0 | 2 | 14 | (Hébrard et al. 2012) |
|  | → HNC + H$_2$ | -276 | 1.2e-11 | 0 | 0 | 3 | 14 |  |
| H + CCN | → C + HCN | -55 | 2.0e-10 | 0 | 0 | 3 | 0 | Ab-initio and RRKM calculations (Bergeat & Loison) in preparation, CNC abundance included in CCN. |
|  | → CNC + H | -20 | 0 |  |  |  |  |  |
| H$^+$ + HCN | → HNC$^+$ + H | -219 | 1.0 | 3.77e-9 | 6.61 | 2 | 0 | Ionpol1 |
| H$^+$ + HNC | → HNC$^+$ + H | -219 | 1.0 | 3.94e-9 | 6.50 | 2 | 0 | Ionpol1 |
|  | → HCN + H$^+$ | -53 | 0 |  |  |  |  |  |
| H$_2$ + CCN$^+$ | → HCNH$^+$ + C | -60 | 9.0e-10 | 0 | 0 | 1.6 | 0 | (Knight *et al.* 1988) |
| H$_2$ + CNC$^+$ | → HCNH$^+$ + C | +46 | 0 |  |  |  |  | Endothermic (Knight *et al.* 1988) |
| C + NH$_2$ | → HCN + H | -554 | 3.0e-11 | -0.20 | -6 | 3 | 0 | (Herbst *et al.* 2000, Talbi 1999a) |
|  | → HNC + H | -499 | 3.0e-11 | -0.20 | -6 | 3 | 0 |  |
| C + HNC | → HCN + C | -53 | 1.6e-10 | 0 | 0 | 3 | 0 | Ab-initio and RRKM calculations (Bergeat & Loison) in preparation. CNC isomer is included into CCN |
|  | → H + CNC | -20 | 0.4e-10 | 0 | 0 | 3 | 0 |  |
| C + H$_2$CN | → H$_2$ + C$_2$N | -276 | 3.0e-11 | 0 | 0 | 4 | 0 | Estimated from (Cho & Andrews 2011, Lau *et al.* 1999, Koput 2003, Osamura & Petrie 2004). |
|  | → HCN + CH | -234 | 3.0e-11 | 0 | 0 | 4 | 0 |  |
|  | → HCCN + H | -257 | 3.0e-11 | 0 | 0 | 4 | 0 |  |
| C + CH$_2$NH | → HCN + CH$_2$ | -264 | 1.0e-10 | 0 | 0 | 2 | 0 | Estimated from (Feng *et al.* 2007) and (Georgievskii & Klippenstein 2005) |
|  | → CH$_2$CN + H | -313 | 1.0e-10 | 0 | 0 | 2 | 0 |  |
| C + CCN | → CN + C$_2$ | -133 | 2.4e-10 | 0 | 0 | 3 | 0 | By comparison with C + NO (Andersson et al. 2003) |
|  | → C$_3$ + N | -103 | 0 |  |  | 0 | 0 |  |
| C + H$_2$CCN | → HC$_3$N + H | -400 | 1.0e-10 | 0 | 0 | 3 | 0 | Rate constant close to capture value, products guessed considering C atom insertion into CH bond |
| C$^+$ + NH$_3$ | → C + NH$_3^+$ | -115 | 0.32 | 1.28e-9 | 3.65 | 2 | 0 | Ionpol1 (Talbi & Herbst 1998) |
|  | → H + HNCH$^+$ | -590 | 0.63 | 1.28e-9 | 3.65 | 2 | 0 |  |
|  | → H + H$_2$NC$^+$ | -371 | 0 | 0 | 0 | 0 | 0 |  |
|  | → H$_2$ + HCN$^+$ | -310 | 0 | 0 | 0 | 0 | 0 |  |
|  | → H$_2$ + HNC$^+$ | -421 | 0.05 | 1.28e-9 | 3.65 | 2 | 0 |  |
| C$^+$ + HCN | → H + CNC$^+$ | -100 | 1.0 | 1.28e-9 | 6.61 | 2 | 0 | Ionpol1 (Clary *et al.* 1990, Clary *et al.* 1985, Anicich *et al.* 1986) |
|  | → H + C$_2$N$^+$ | +6 | 0 | 0 | 0 | 0 | 0 |  |
| C$^+$ + HNC | → H + CNC$^+$ | -155 | 0.5 | 1.34e-9 | 6.5 | 2 | 0 | Ionpol1 |
|  | → H + CCN$^+$ | -49 | 0.5 | 1.34e-9 | 6.5 | 2 | 0 |  |
| C$^+$ + CCN | → CCN$^+$ + C | -49 | 0.3 | 1.30e-9 | 0.62 | 2 | 10 | Thermochemsitry for CCN and CNC (and ions) from (Mebel & Kaiser 2002, Harland & McIntosh 1985), DFT calculations showing no barrier for CCCN$^+$ and CCNC$^+$ formation. 10-30K: Ionpol1, 40-300K: Ionpol2 |
|  | → CNC$^+$ + C | -155 | 0.4 | 1.30e-9 | 0.62 | 2 | 10 |  |
|  | → CN + C$_2^+$ | -81 | 0.3 | 1.30e-9 | 0.62 | 2 | 10 |  |
|  | → CN$^+$ + C$_2$ | -7 | 0 |  |  |  |  |  |
| C$^+$ + H$_2$CCN | → C + CH$_2$CN$^+$ | -114 | 1.0 | 1.61e-9 | 5.81 | 2 | 0 | Ionpol1 |
| N + CH | → CN + H | -416 | 1.4e-10 | 0.42 | 0 | 1.6 | 7 | (Daranlot *et al.* 2013) |
| N + CH$_2$ | → HCN + H | -510 | 5.0e-11 | 0.17 | 0 | 3 | 0 | (Herbst et al. 2000, Hébrard et al. 2012) |
|  | → HNC + H | -455 | 3.0e-11 | 0.17 | 0 | 3 | 0 |  |
| N + CH$_3$ | → H$_2$CN + H | -153 | 5.6e-11 | 0.17 | 0 | 1.6 | 7 | (Marston et al. 1989a, Cimas & Largo 2006) (Nguyen *et al.* 1996, Hébrard *et al.* 2012) |
|  | → HCN + H + H | -48 | 0.6e-11 | 0.17 | 0 | 2 | 7 |  |
| N + CH$_3^+$ | → H + HNCH$^+$ | -401 | 0 |  |  |  |  | (Scott *et al.* 1999, Talbi 1999b). The isomers produced are non-determined but it is not critical. |
|  | → H + H$_2$NC$^+$ | -182 | 6.1e-11 | 0 | 0 | 2 | 10 |  |
|  | → H$_2$ + HCN$^+$ | -121 | 3.3e-11 | 0 | 0 | 2 | 0 |  |
|  | → H$_2$ + HNC$^+$ | -232 | 0 |  |  |  | 10 |  |
|  |  |  |  |  |  |  | 0 |  |
| N + C$_2$H | → $^2$CCN + H | -132 | 1.5e-10 | 0.17 | 0 | 2 | 0 | Ab-initio and RRKM calculations (Bergeat & Loison) in preparation |
|  | → HCN + C | -187 | 8.0e-12 | 0.17 | 0 | 2 | 0 |  |
|  | → HNC + C | -132 | 2.0e-12 | 0.17 | 0 | 2 | 0 |  |
| N + C$_2$H$_3$ | → H$_2$CCN + H | -304 | 6.4e-11 | 0.17 | 0 | 2 | 7 | (Payne *et al.* 1996) |
|  | → C$_2$H$_2$ + NH | -193 | 1.3e-11 | 0.17 | 0 | 2 | 7 |  |
| N + C$_2$H$_3^+$ | → HC$_2$N$^+$ + H$_2$ |  | 2.2e-10 | 0 | 0 | 3.0 | 0 | (Federer *et al.* 1986) |
| N + C$_2$H$_4^+$ | → CH$_2$CNH$^+$ + H |  | 3.0e-10 | 0 | 0 | 1.8 | 0 | (Scott *et al.* 1999) |
| N + C$_2$H$_5$ | → H$_2$CN + CH$_3$ | -200 | 6.0e-11 | 0.17 | 0 | 2 | 7 | (Stief *et al.* 1995, Yang *et al.* 2005b) |
|  | → C$_2$H$_4$ + NH | -186 | 4.0e-11 | 0.17 | 0 | 3 | 21 |  |
| N + H$_2$CN | → HCN + NH | -229 | 5.0e-12 | 0 | 0 | 4 | 0 | (Hébrard et al. 2012). |
|  | → CH$_2$ + N$_2$ | -331 | 4.0e-11 | 0 | 0 | 3 | 0 |  |
| N + CCN | → CN + CN | -216 | 9.0e-11 | 0.17 | 0 | 3 | 0 |  |

| Reactants | Products | | | | | | | Comments |
|---|---|---|---|---|---|---|---|---|
| N + H$_2$CCN | → H$_2$CN + CN<br>→ HCN + HNC<br>→ H + HC$_2$N$_2$<br>→ H + H + C$_2$N$_2$ | -45<br>-403<br>-77<br>+17 | 6.0e-11<br>0<br>0<br>0 | 0.17 | 0 | 3 | 0 | By comparison with N + C$_2$H$_5$ (Stief *et al.* 1995, Yang *et al.* 2005b). The H + HC$_2$N$_2$ exit channel is likely a minor one's (there is a small exit barrier (Basiuk & Kobayashi 2004)) but HC$_2$N$_2$ is not in our model. |
| O + H$_2$CN | → OH + HCN<br>→ OH + HNC<br>→ H + HCNO | -335<br>-280<br>-105 | 4.0e-11<br>1.0e-11<br>0 | 0<br>0 | 0<br>0 | 3<br>3 | 0<br>0 | Estimation using (Fikri *et al.* 2001, Zhang *et al.* 2004). HCNO is likely produced but is not in our model. |
| O + CCN | → $^1$CO + $^2$CN | -614 | 7.0e-11 | 0.17 | 0 | 3 | 0 | Capture rate constant considering no barrier |
| O + H$_2$CCN | → H$_2$CO + CN<br>→ HCO + HCN<br>→ H + CO+HCN | -178<br>-326<br>-262 | 1.0e-11<br>1.0e-11<br>2.0e-11 | 0<br>0<br>0 | 0<br>0<br>0 | 3<br>3<br>3 | 10<br>10<br>10 | Rate constant deduced from (Hoyermann & Seeba 1995), branching ratio estimated using exothermicities. |
| CH + NH | → HCN + H<br>→ HNC + H | -599<br>-546 | 5.0e-11<br>5.0e-11 | 0<br>0 | 0<br>0 | 3<br>3 | 0<br>0 | (Takahashi & Takayanagi 2007, Georgievskii & Klippenstein 2005) |
| CH + NH$_3$ | → CH$_2$NH + H | -240 | 1.69e-10 | -0.56 | 28 | 1.6 | 0 | (Bocherel *et al.* 1996, Blitz *et al.* 2012, Zabarnick *et al.* 1989) |
| CH + HCN | → CCN + H$_2$<br>→ HCCN + H | -42<br>-23 | 1.4e-10 | 0 | 0 | 2 | 10<br>0 | (Zabarnick *et al.* 1991, Du & Zhang 2006, Osamura & Petrie 2004) |
| CH + HNC | → CCN + H$_2$<br>→ HCCN + H | -97<br>-76 | 1.4e-10 | 0 | 0 | 2 | 10<br>0 | (Hébrard et al. 2012) |
| CH$_2$ + NH | → H$_2$CN + H<br>→ HCN + H + H<br>→ HNC + H$_2$ | -280<br>-175<br>-558 | 3.0e-11<br>3.0e-11<br>5.0e-12 | 0<br>0<br>0 | 0<br>0<br>0 | 3<br>3<br>4 | 21<br>21<br>21 | By comparison with H$_2$CNH decomposition (Zhou & Schlegel 2009, Nguyen *et al.* 1996) (Georgievskii & Klippenstein 2005) |
| CH$_3$ + NH | → CH$_2$NH + H<br>→ N + CH$_4$ | -192<br>-104 | 1.3e-10<br>0 | 0.17 | 0 | 2 | 10<br>0 | (Redondo *et al.* 2006, Georgievskii & Klippenstein 2005) |
| CH$_3^+$ + HCN | → CH$_3$CNH$^+$ +hv | | 9.0e-9 | -0.5 | 0 | 10 | 0 | (McEwan *et al.* 1980, Bass *et al.* 1981, Bates 1983, Herbst 1985) |
| CH$_3^+$ + HNC | → CH$_3$CNH$^+$ +hv | | 9.0e-9 | -0.5 | 0 | 10 | 0 | = HCN + CH$_3^+$ |
| C$_2$ + HCN | → C$_3$N + H | -148 | 2.0e-10 | 0 | 0 | 3 | 0 | (Silva *et al.* 2009, Georgievskii & Klippenstein 2005) |
| C$_2$ + HNC | → C$_3$N + H | -201 | 2.0e-10 | 0 | 0 | 4 | 0 | (Silva et al. 2009, Georgievskii & Klippenstein 2005) |
| C$_2$H + HNC | → HC$_3$N + H | -182 | 2.0e-10 | 0 | 0 | 3 | 0 | (Fukuzawa & Osamura 1997, Petrie 2002, Georgievskii & Klippenstein 2005) |
| CN + CH$_3$ | → H$_2$CCN + H | -108 | 1.0e-10 | 0 | 0 | 3 | 21 | (Yang *et al.* 2005a, Georgievskii & Klippenstein 2005) |
| CN + CH$_3^+$ | → CH$_2$CN$^+$ + H | -63 | 1.0 | 1.29e-9 | 2.84 e0 | 2 | 0 | KIDA, Ionpol1 |
| C$_3$N + HNC | → NC$_4$N + H | -127 | 2.0e-10 | 0 | 0 | 3 | 0 | (Petrie & Osamura 2004, Georgievskii & Klippenstein 2005) |
| HCNH$^+$ + e$^-$ | → HCN + H<br>→ HNC + H<br>→ CN + H +H | -596<br>-543<br>-78 | 9.38e-8<br>9.38e-8<br>9.24e-8 | -0.65<br>-0.65<br>-0.65 | 0<br>0<br>0 | 2<br>2<br>2 | 10<br>0<br>100<br>100 | (Semaniak et al. 2001) HNC/HCN ratio from (Mendes et al. 2012, Barger et al. 2003) |
| H$_2$NC$^+$ + e$^-$ | → HCN + H<br>→HNC + H<br>→ CN + H + H | -815<br>-762<br>-297 | 6.6e-8<br>6.6e-8<br>6.5e-8 | -0.5<br>-0.5<br>-0.5 | 0<br>0<br>0 | 3<br>3<br>3 | 10<br>0<br>100<br>100 | The energized HNC** produced should follow the same behaviour than the HCN**/HNC** produced through HCNH$^+$ + e$^-$. |
| CH$_2$CN$^+$ + e$^-$ | → HCCN + H<br>→ CCN + H + H<br>→ CN + CH$_2$<br>→ CH + HCN | -493<br>-76<br>-380<br>-470 | 3.0e-7<br>2.0e-7<br>2.0e-7<br>2.0e-7 | -0.5<br>-0.5<br>-0.5<br>-0.5 | 0<br>0<br>0<br>0 | 3<br>3<br>3<br>3 | 0<br>0<br>0<br>0 | By comparison with similar reactions (Mitchell et al. 1986). Branching ratios roughly deduced from (Plessis *et al.* 2012). |
| CH$_3$CN$^+$ + e$^-$ | → H$_2$CCN + H<br>→ HCCN +H+H<br>→ CN + CH$_3$<br>→ CH$_2$ + HCN<br>→ H$_2$ + H + CCN | -778<br>-325<br>-670<br>-730<br>-345 | 2.0e-7<br>3.0e-7<br>1.0e-7<br>1.0e-7<br>2.0e-7 | -0.5<br>-0.5<br>-0.5<br>-0.5<br>-0.5 | 0<br>0<br>0<br>0<br>0 | 3<br>3<br>3<br>3<br>3 | 0<br>0<br>0<br>0<br>0 | By comparison with similar reactions (Mitchell et al. 1986). Branching ratios roughly deduced from (Plessis *et al.* 2012). |
| CH$_3$CNH$^+$ + e$^-$ | → CH$_3$CN + H<br>→ CH$_2$CN +H+H<br>→ CH$_3$ + HNC<br>→ CH$_3$ + HCN | -533<br>-134<br>-544<br>-489 | 1.3e-7<br>8.0e-8<br>6.0e-8<br>6.0e-8 | -0.5<br>-0.5<br>-0.5<br>-0.5 | 0<br>0<br>0<br>0 | 3<br>3<br>3<br>3 | 0<br>0<br>0<br>0 | (Mitchell et al. 1986). Branching ratios roughly deduced from (Plessis *et al.* 2012). |
| HC$_3$NH$^+$ + e$^-$ | → C$_2$H + HNC<br>→ C$_2$H + HCN<br>→ H + HCNCC<br>→ H + HCCNC<br>→ H + HC$_3$N<br>→ H + HNCCC | -377<br>-432<br><br><br>-559<br> | 3.75e-8<br>3.75e-8<br>1.2e-8<br>1.2e-8<br>1.5e-7<br>8.5e-8 | -0.5<br>-0.5<br>-0.5<br>-0.5<br>-0.5<br>-0.5 | 0<br>0<br>0<br>0<br>0<br>0 | 2<br>2<br>2<br>2<br>2<br>2 | 0<br>0<br>0<br>0<br>0<br>0 | New HCN channel (Osamura *et al.* 1999) |

**Table 3: Molecular abundances (relative to $H_2$) observed in TMC-1 and L134N.**

Notes: a(b) refers to a×10$^b$. Unless otherwise indicated abundances correspond to the positions TMC-1 $\alpha_{J2000}$ = 04$^h$ 41$^m$ 41$^s$.88, $\delta_{J2000}$ = +25° 41$^m$ 27$^s$ (cyanopolyyne peak) and L134N $\alpha_{J2000}$ = 15$^h$ 54$^m$ 06$^s$.55, $\delta_{J2000}$ = -2° 52$^m$ 19$^s$. Most abundances were derived from observed column densities adopting N($H_2$) = 10$^{22}$ cm$^{-2}$ for both TMC-1 and L134N (Goldsmith *et al.* 2007). HCN and HNC abundances do not take into consideration the new rotational excitation cross sections (Sarrasin et al. 2010, Dumouchel et al. 2010, Dumouchel et al. 2011) and are determined using H$^{13}$CN and HN$^{13}$C with a $^{12}$C/$^{13}$C ratio equal to 68 (Milam et al. 2005). Values denoted by * have been deduced from an LTE analysis considering optically thin absorption.

|  | TMC-1 |  | L134N (L183) |  |
|---|---|---|---|---|
| CN | 7.4(-10) | (Pratap et al. 1997) | 4.8(-10) | (Dickens et al. 2000) |
|  | 2.9(-8)* | (Crutcher *et al.* 1984) | 8.0(-10)* | (Hily-Blant *et al.* 2010) |
| HCN | 1.1(-8) | (Pratap et al. 1997) | 5.1(-9)* | (Swade 1989) |
|  | 7.1(-9) | (Hirota et al. 1998) | 7.3(-9) | (Dickens et al. 2000) |
|  | 3.4(-8)* | (Irvine & Schloerb 1984) | 6.3(-9) | (Hirota et al. 1998) |
|  |  |  | 3.4(-10)* | (Hily-Blant et al. 2010) |
| HNC | 2.6(-8) | (Pratap et al. 1997) | 1.2(-8)* | (Swade 1989) |
|  | 1.2(-8) | (Hirota et al. 1998) | 2.6(-8) | (Dickens et al. 2000) |
|  | 5.4(-8)* | (Irvine & Schloerb 1984) | <7.6(-9) | (Hirota et al. 1998) |
|  |  |  | 9.0(-10)* | (Hily-Blant et al. 2010) |
| H$_2$CN | 1.5(-11)* | (Ohishi et al. 1994) | - |  |
| CH$_3$CN | 4.5(-10) | (Turner *et al.* 1999) | - |  |
|  | 6.0(-10)* | (Ohishi & Kaifu 1998) | - |  |
| H$_2$CCN | 5.0(-9)*$^{(?)}$ | (Irvine & et al. 1988) | - |  |
| HC$_3$N | 1.6(-8)* | (Takano *et al.* 1998) | - |  |
|  | 4.2(-9) | (Pratap et al. 1997) | - |  |
|  | 6.0(-9) *$^{(?)}$ | (Ohishi *et al.* 1992) | 1.6(-10)*$^{(?)}$ | (Ohishi *et al.* 1992) |
|  |  |  | 4.3(-10) | (Dickens et al. 2000) |
| HCNH$^+$ | 1.9(-9)* | (Schilke et al. 1991) | <3.1(-9) | (Ohishi *et al.* 1992) |

**Figure 1**: Potential energy curves for HCN-HNC neutral and ions as a function of the angle between H and the CN center of mass. The energies were calculated at the MRCI+Q/aug-cc-pVQZ level using Molpro, with all inter-nuclear distances being optimized for each angle. Lines correspond to the ZPE levels. Values in parentheses are the relative energies with respect to the HCN ground state in kJ/mol including ZPE (at the MRCI+Q/aug-cc-pVQZ level).

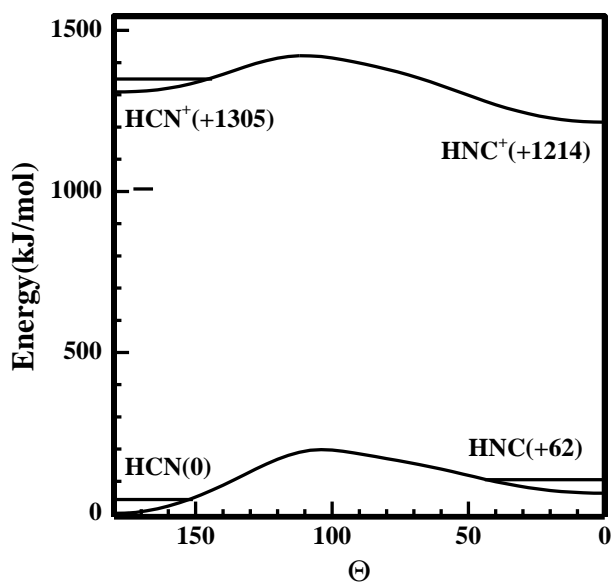

**Figure 2**: HCN/HNC abundance ratios as a function of time (upper panels) predicted by the model using the old (left plot) and new (right plot) rate constants. The middle panels represent the predicted HCN (solid curve) and HNC (dashed curve) abundances as a function of time for the old (left plot) and new (right plot) rate constants. Horizontal lines represent the mean observed abundances in dense clouds (see text). The lower panels represent the predicted main atoms and molecules (grey represent old network and black new network).

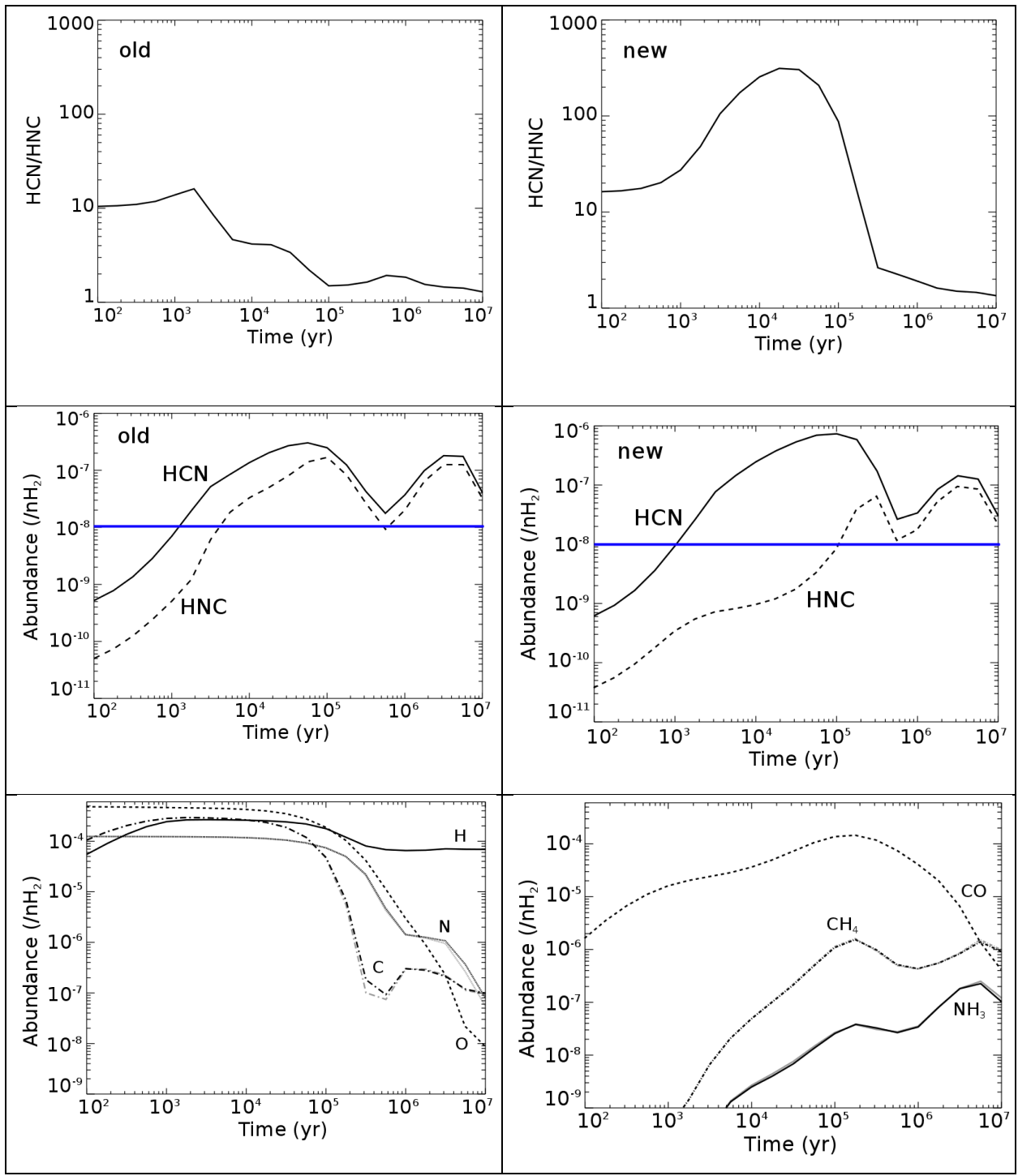

**Figure 3**: Relative abundances of CN, HCNH+, CCN, $H_2CN$, $HC_3N$ and $CH_3CN$ (versus $H_2$) as a function of time obtained with the old (dashed curves) and new (solid curves) rate constants. Horizontal lines represent the mean observed abundances in dense clouds (see text).

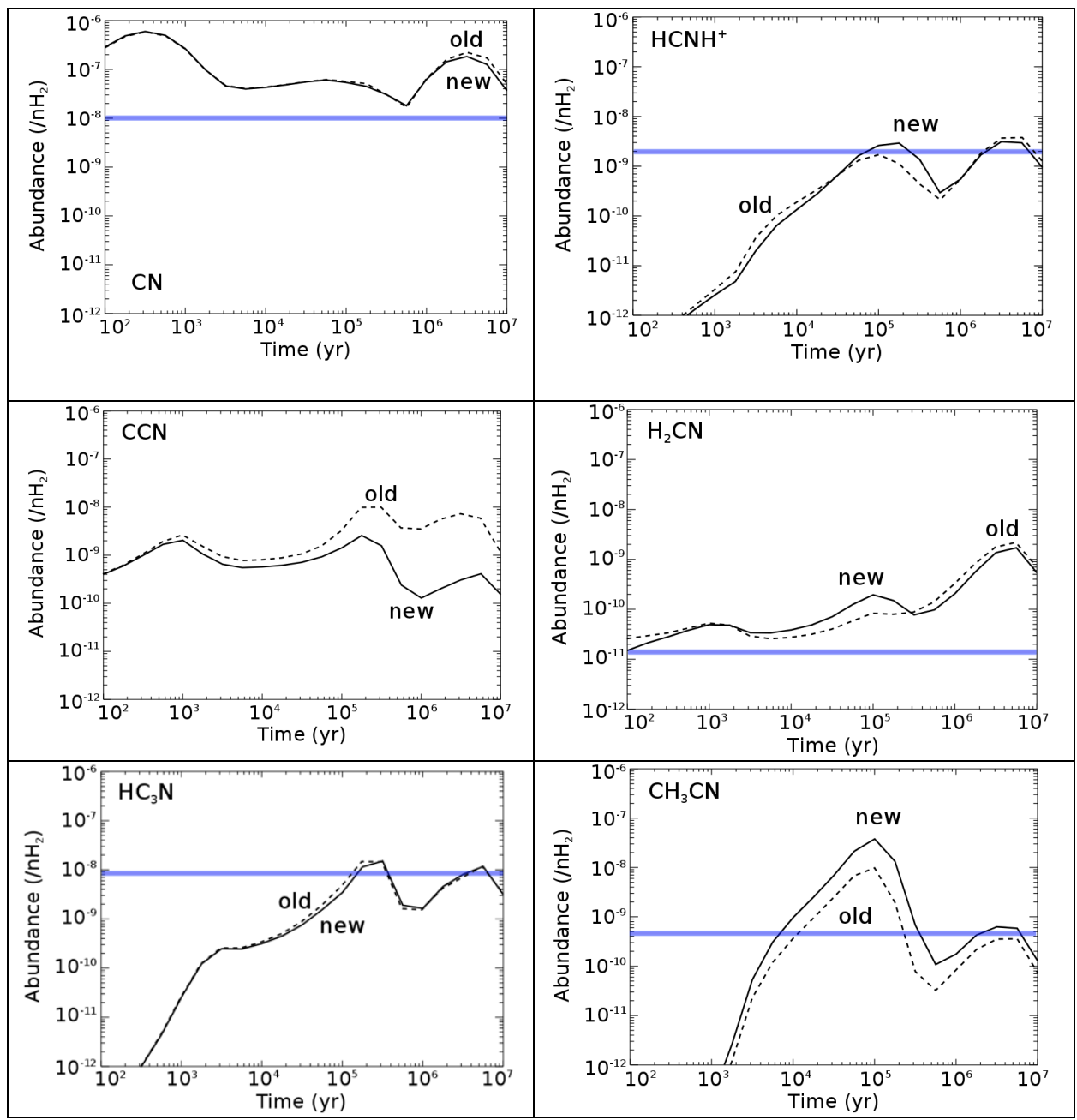

**Figure 4**: Fluxes for the main HCN and HNC production and destruction reactions as a function of the time with the new rate constants.

Top left plot: (1) N + CH$_2$, (2) HCNH$^+$ + e$^-$ → HCN + H, (3) H + CCN, and (4) C + HNC.

Top right plot: (1) N + CH$_2$ and (2) HCNH$^+$ + e$^-$ → HNC + H.

Bottom left plot: (1) C$^+$ + HCN, (2) sum of HCN + H$^+$ and HCN + He$^+$, (3) HCN + H$_3^+$, HCN + H$_3$O$^+$ and HCN + HCO$^+$, (4) CH$_3^+$ + HCN, and (5) sticking of HCN on the grains.

Bottom right plot: (1) C$^+$ + HNC, (2) C + HNC, (3) sum of HNC + H$_3^+$, HNC + H$_3$O$^+$ and HNC + HCO$^+$, and (4) sum of HNC + H$^+$ and HNC + He$^+$.

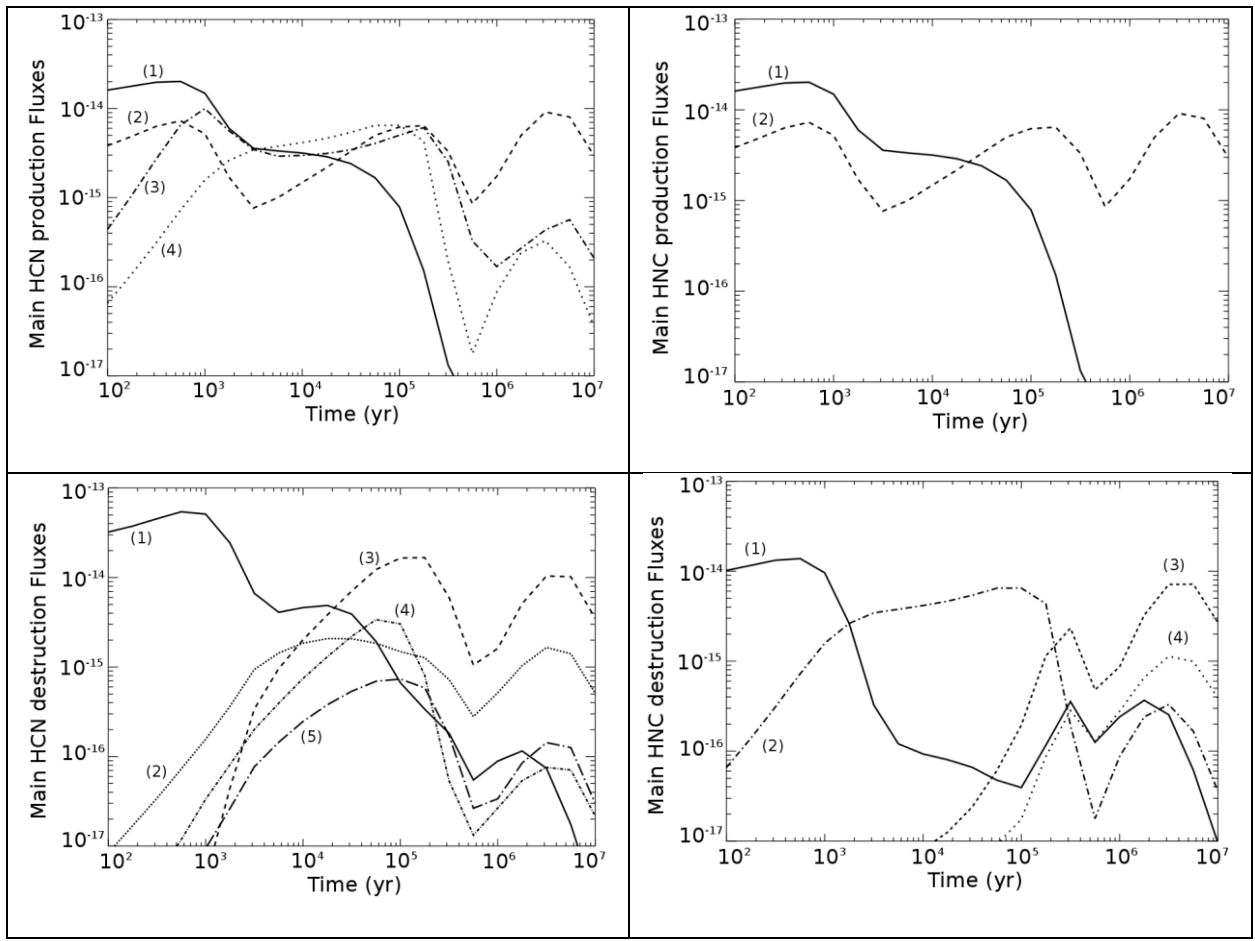

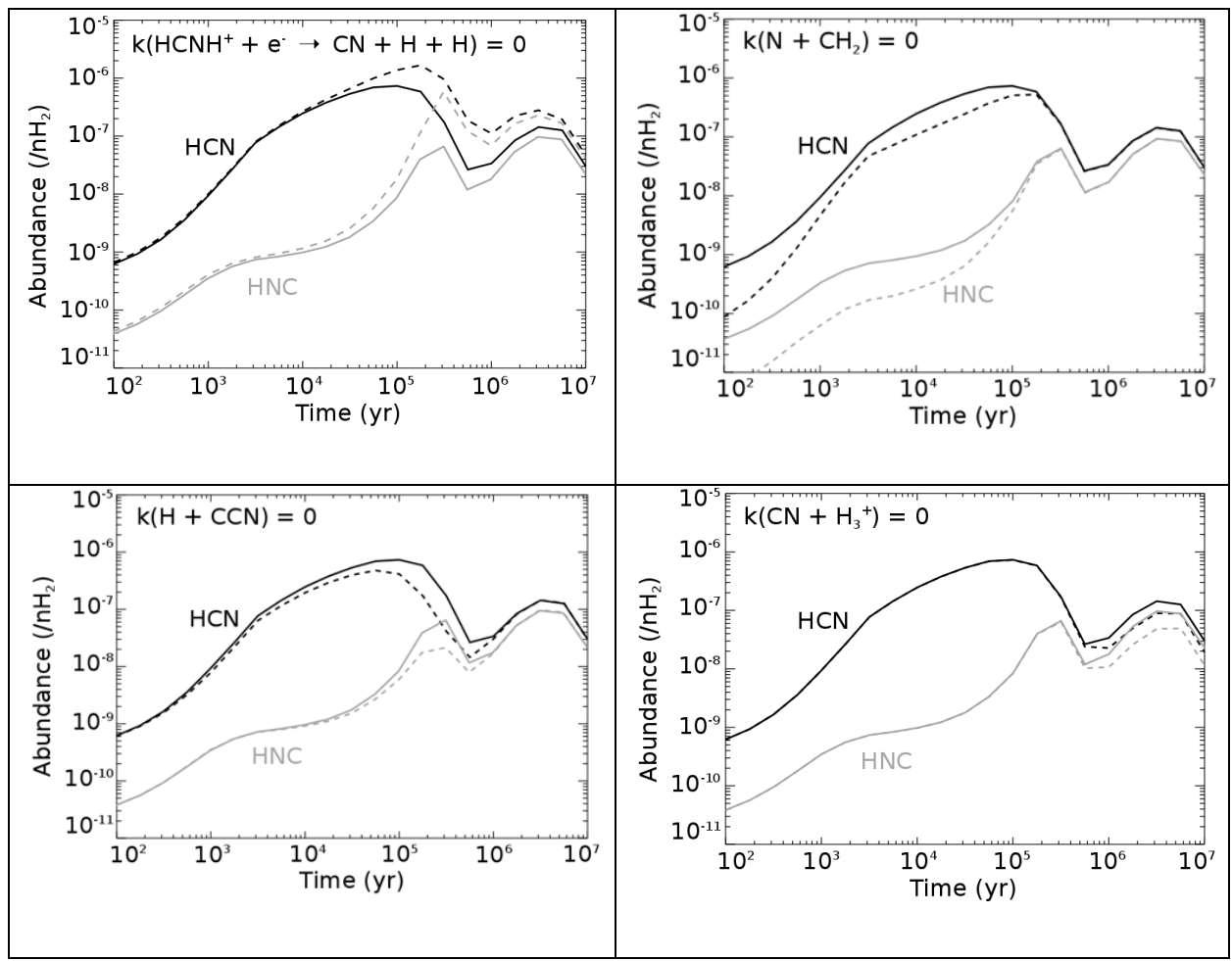

**Figure 5**: Relative abundances of HCN (black curves) and HNC (grey curves) (versus $H_2$) for the new rate constants. The dashed curves in each plot have been obtained with the new rate constants but one of the reactions has been set to zero as indicated.

MOLPRO, version 2009.1, a package of ab initio programs, H.-J.Werner, P. J. Knowles, R. Lindh, F. R. Manby, M. Schütz, P. Celani, T. Korona, A. Mitrushenkov, G. Rauhut, T. B. Adler, R. D.Amos, A. Bernhardsson, A. Berning, D. L. Cooper, M. J. O. Deegan, A. J. Dobbyn, F. Eckert, E. Goll, C. Hampel, G. Hetzer, T. Hrenar, G. Knizia, C. Köppl, Y. Liu, A. W. Lloyd, R. A.Mata, A. J. May, S. J. McNicholas, W. Meyer, M. E. Mura, A. Nicklass, P. Palmieri, K. Pflüger, R. Pitzer, M. Reiher, U. Schumann, H. Stoll, A. J. Stone, R. Tarroni, T. Thorsteinsson, M.Wang, and A. Wolf, see http://www.molpro.net.